 \documentclass[twocolumn,nofootinbib,showkeys,superscriptaddress]{revtex4-1} 

\usepackage{amsmath}
\usepackage{graphicx}
\usepackage{dcolumn}
\usepackage{bm}
\usepackage{color}
\usepackage[normalem]{ulem}
\usepackage[colorinlistoftodos]{todonotes}
\usepackage[colorlinks=true, allcolors=blue]{hyperref}
\usepackage{epstopdf}
\usepackage[english]{babel}
\usepackage[utf8x]{inputenc}
\usepackage[T1]{fontenc}

\usepackage{tikz}

\begin{document}

\title{How the geometry of cities explains urban scaling laws and determines their exponents}

\author{Carlos Molinero}
\email[Corresponding author: ]{molinero@csh.ac.at}
\affiliation{Complexity Science Hub Vienna, Josefst{\"a}dterstrasse 39, A-1080 Vienna, Austria}
\affiliation{Austrian Institute of Technology, Giefinggasse 2, A-1210 Vienna, Austria}
\affiliation{Centre for Advanced Spatial Analysis. University College of London, 90 Tottenham Court Road, W1T 4TJ London, UK}

\author{Stefan Thurner}
\affiliation{Complexity Science Hub Vienna, Josefst{\"a}dterstrasse 39, A-1090 Vienna, Austria}
\affiliation{Section for the Science of Complex Systems, CeMSIIS, Medical University of Vienna, Spitalgasse 23, A-1090, Vienna, Austria}
\affiliation{Santa Fe Institute, 1399 Hyde Park Road, Santa Fe, NM 87501, USA}
\affiliation{IIASA, Schlossplatz 1, 2361 Laxenburg, Austria}

\date{Version \today}

\begin{abstract}

  Urban scaling laws relate socio-economic, behavioral, and physical variables to the population size of cities and allow for a new paradigm of city planning, and an understanding of urban resilience and economies. Independently of culture and climate, almost all cities exhibit two fundamental scaling exponents, one sub-linear and one super-linear that are related. Here we show that based on fundamental fractal geometric relations of cities we derive both exponents and their relation. Sub-linear scaling arises as the ratio of the fractal dimensions of the road network and the distribution of the population in 3D. Super-linear scaling emerges from human interactions that are constrained by the city geometry. We demonstrate the validity of the framework with data on 4750 European cities.
  We make several testable predictions, including the relation of average height of cities with population size, and that at a critical population size, growth changes from horizontal densification to three-dimensional growth.

\end{abstract}


\maketitle

In the past decade a ``science of cities'' \cite{batty2013new,west2017scale,barthelemy2016structure} emerged as a new discipline that tries to extract useful knowledge from the
vast datasets on cities that are now available \cite{pumain2004scaling,fuller2009scaling,bettencourt2010urban,kuhnert2006scaling,gastner2006optimal}.
%
One of the surprising findings is that many of the hundreds of quantities and variables that characterize the
dynamics, functioning, and performance of a city, exhibit power law relations.
This means that these quantities, $X$, are related to other quantities, $Y$, in a particularly simple way,
\begin{equation}
X \sim Y^{\gamma} \quad ,
\end{equation}
where $\gamma$ is the scaling exponent. Let $Y=p$ denote the population size of a city.
Obviously, several quantities scale linearly ($\gamma=1$) with population, such as water consumption,
housing, or the number of employments \cite{bettencourt2007growth}.
However, non-trivial urban scaling laws abound and appear in a vast number of different contexts.
For example, scaling laws with respect to population size were found for
$X=$ GDP \cite{bettencourt2010urban,strano2016rich},
number of patents \cite{bettencourt2007invention},
walking speed \cite{noulas2012tale}, or
crime rates \cite{bettencourt2010urban}.
The associated scaling exponent for these relations appears to be in a range of $\gamma  \sim 1.1 - 1.2$.
Since $\gamma > 1$, it is sometimes referred to as the super-linear scaling exponent, $\gamma_{\rm sup}$.
For other quantities, such as
$X=$ total length of the road network \cite{samaniego2008cities},
length of electrical cables \cite{bettencourt2007growth},
number of facility locations \cite{gastner2006optimal},
or petrol stations \cite{kuhnert2006scaling},
the associated scaling exponent is often found in a range of $\gamma_{\rm sub} \sim 0.8 - 0.9$, and is called sub-linear scaling\footnote{In \cite{gastner2006optimal} they find an exponent of $0.66$, however, with respect to another variable. This variable (density)
approximately scales with $0.15$ with respect to population, so that effectively they have an exponent of $0.15+0.66=0.81$.
}.
%
Note that technically it is all but trivial to quantify urban scaling exponents reliably and consistently and some works question the measurement techniques used in a large fraction of the literature \cite{leitao2016scaling,shalizi2011scaling}.
A major difficulty is a proper definition of city boundaries, which is at the heart of some discrepancies in several works \cite{arcaute2015constructing,cottineau2017diverse,louf2014scaling}.
Depending on the notion of city boundaries, it has been shown that exponents for a system can vary substantially, sometimes even
from a sub-linear to a super-linear behaviour.
To avoid this issue, we propose an approach to obtain city boundaries directly from population data; for details see SI.

Urban scaling is of immediate practical relevance for a number of reasons.
First, they allow us to compute the detailed economies of scale in cities.
They relate the size of cities to efficiency gains or losses for a wide range of quantities that determine life in cities.
For example, if a quantity like the total length of the road network
scales sub-linearly with population size, this means that the cost per person
decreases with city size; the larger a city becomes the more efficient it will be with respect to this variable.
If the population of city A is $x$ times larger than B, sub-linear scaling means that the per capita effort in city
A is a factor $x^{\gamma-1}<1$  less than in B.
Second, to compare cities of different population size, it is necessary to correctly rescale the respective quantities before the comparison.
If one would directly compare, for example, the per capita GDP of a large and a small city, due to super-linear scaling, the large city will
have a bias toward larger GDP values that is only due to scaling, and not to e.g. better management of the large city.
Third, since urban scaling laws appear to be 
largely similar across countries and cultures, they can be used for urban planning,
in particular for anticipating consequences of rapid growth.
If a city is expected to double in size within the next decades, depending on the scaling exponents,
dozens of performance indicators, growth rates, infrastructure costs, etc. can be inferred and used proactively in city planning.
Given a level of growth, scaling laws pose clear constraints to urban performance indicators and possibilities for change.

Urban scaling laws exhibit two remarkable phenomena.
The first is that even though cities can be very different, as we know from everyday experience,
urban power laws and their exponents are not. Similar exponents have been reported across countries, regions, and continents,
and have been called universal \cite{bettencourt2010unified}. The extent of this universality is currently under debate since other studies have shown that measured exponents differ between countries and depend on the way cities are defined \cite{arcaute2015constructing}.
The second is that super- and sub-linear scaling exponents tend to add up to two \cite{ribeiro2017model},
$\gamma_{\rm sup} + \gamma_{\rm sub} = 2$.

\begin{figure*}[t]
\centering
\includegraphics[width=0.9\textwidth]{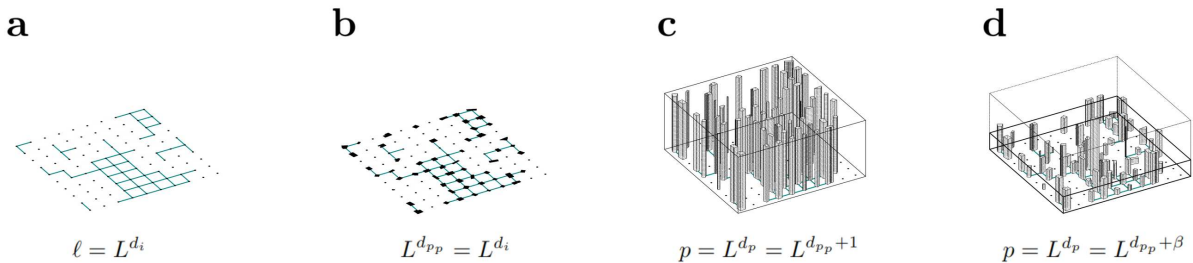}
\caption{(a) Street network in a section of the city of size $L$.
	The length of the street network with fractal dimension $d_i$ expands with the linear scale $L$ as $\ell=L^{d_i}$.
	(b) Buildings are located along the street network and are attached to it. Since people live and work mostly in buildings,
	the fractal dimension of the ``projected population'' (the actual population fractal projected onto the 2D surface, where streets are embedded)
	should have a similar fractal dimension $d_{p_p}=d_i$.
	(c) If all buildings had the same height, the fractal dimension of the population, $d_p$, should be the projected population dimension plus 1,
	$d_p=d_{p_p}+1$.
	(d) More realistically, not all buildings have the same height, and the fractal dimension of the populations is $d_p=d_{p_p}+\beta$,
	where $\beta$ captures the fractal dimension along the third dimension.
\label{fig::fractalPop}}
\end{figure*}

Until today, a general understanding of urban scaling laws is still under debate.
In particular, the origin of the values of the super- and sub-linear exponents,
and
why they cluster in specific ranges
and the addition law, call for a coherent and comprehensive explanation.
Are the observed power laws of statistical origin \cite{thurnerhanelklimek2018}, or do they arise from deep underlying behavioral or geometrical rules?
If the latter is true, how can geometry be used to learn how cities work, and how to evade the constraints on growth and change, such that cities become able to adapt and meet the challenges of the coming decades?. First steps taken in this direction were proposed in \cite{bettencourt2013origins}.

Various explanations have been suggested for the emergence of scaling in the urban context.
Some use underlying network structures of the social tissue. In \cite{arbesman2009superlinear} the authors focus on the social network structure of cities understood as a hierarchical tree. This allows them to define a distance in the tree that will be used to calculate the probability of people interacting, to calculate the overall productivity of the city as proportional to the number of interactions.
They are able to reproduce the super-linear exponent ranges, but the approach is highly theoretical and uses several assumptions that cannot be tested, such as the structure of the social ties (tree-like) or the shape of the decay of interactions with the distance in that topological structure.
In \cite{yakubo2014superlinear} the authors use a geographical network embedded in a fractal Euclidean space,
which is proposed as an explanation of sub- and super-linear exponents. However, they need to create two parameters that are difficult to measure further complicating the model, such as the attractiveness of a person or the exponents that drive the decay of interactions with respect to the distance in their model.
Another approach is based on path-dependent evolution of innovations \cite{pumain2006evolutionary},
where cumulative cycles of innovation give rise to the growth of cities, which further reinforce the next set of innovations. The authors give a longitudinal explanation of scaling exponents that depend on the cycle of innovation of each sector, relegating more mature technologies to smaller cities while new products are generated in the largest cities. This explanation of economic innovation cycles does not explain other scaling exponents that relate to physical quantities such as the scaling of infrastructure or of the location of gas stations.

Two recent works propose to explain the observed exponents partly on the basis of the underlying geometrical structure of cities.
In \cite{bettencourt2013origins} authors consider growth models for cities in which an equilibrium between costs and benefits produce the scaling exponents and assume that cities are space-filling fractals. Most cities will have a fractal dimension lower than 2 since in every settlements there exists empty spaces and voids of different sizes, such as parks and open public spaces, which leads to measured fractal dimensions that fall
consistently in the range $d_i\in [1.2 - 1.93]$, depending on the city \cite{murcio2015multifractal,batty1994fractal,frankhauser1992fractal,frankhauser1990aspects,batty1987urban,batty1987fractal}.
The model of \cite{ribeiro2017model} builds on the notion that interactions between people decay with distance in a specific way, and
assume that the fractal dimension of the population, $d_p$, is equal to that of the infrastructure. In particular  they expect it to be around
$d_p \sim1.7$. However, given that the population lives in three-dimensional buildings, its fractal dimension should be expected to be
definitively larger than $d_i$, typically also larger than 2.
Both models use geometric arguments, but do not attempt to directly relate the geometry of a city to the observed scaling exponents.

This is exactly what we propose in this work. Scaling laws can often be explained directly from the geometry of the underlying
structures of a system. Classic examples include Galileo's understanding of the relation between the shape of animals and their body mass \cite{schmidt1975scaling}, and the understanding of the allometric scaling laws in biology on the basis of the fractal geometry
of the branching of vascular systems \cite{west1997general, west1999fourth, west2001general}.
In the same spirit we provide a simple and a direct geometrical explanation of urban scaling exponents, derived from the fractal geometry of cities.
The problem is challenging since cities across countries, latitudes, and cultures are different and so is their geometry.
How should cities that are significantly different in their geometry lead to similar scaling exponents?
The basic idea is that we focus on the ratio of two geometric aspects of a city, the fractal dimension of its infrastructure (street networks)
\cite{batty1989urban,Batty_LongleyFractasl1994,batty2008size,frankhauser1998fractal},
and the fractal dimension of the population, meaning the dimension of the object that represents the spatial distribution of the population in a city.
The fractal of the population can be imagined as the cloud of people that is obtained by identifying the position of every
person in three dimensions.
The corresponding sub-linear exponent turns out to be the ratio of the two dimensions.
It has been suggested that the super-linear exponent is related to the number of possible interactions of people in a city \cite{bettencourt2013origins}.
In our framework, the super-linear exponent is a direct consequence of the number of possible interactions between people that share a common location,
which can be explained in terms of the geometric configuration of a city.

Within this geometric framework we are able not only to understand the origin of the specific super- and sub-linear scaling exponents in a new light and
why they add up to 2, we can also predict a number of geometric scaling laws, such as
the average height of a city,
the length of the road network,
the area that contains a city, and
the number of interactions;
all as a function of its population. All these predictions are confirmed empirically to a large level of precision.

\section*{Results}

\begin{figure*}[t]
\centering
\includegraphics[width=1\textwidth]{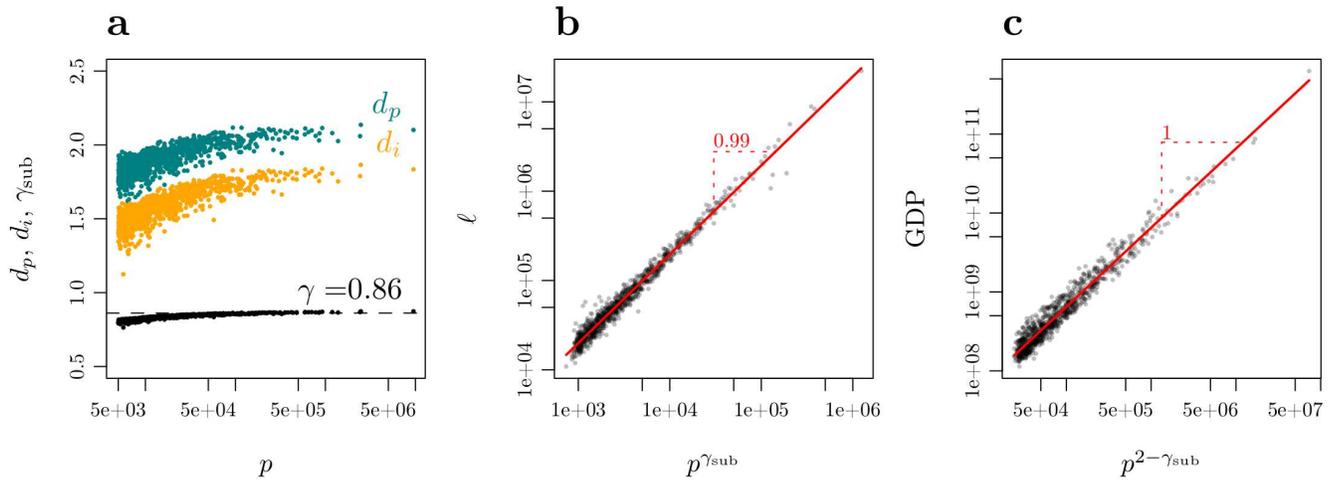}
\caption{(a) Fractal dimensions of the street network, $d_i$, (orange) and for the population, $d_p$, (blue)
		for 1000 cities in the UK as a function of their population size $p$. While the fractal dimensions are strongly
		size dependent, their ratio, $\gamma_{\rm sub}=d_i/d_p$ (black), is not. It is found to be approximately
		constant, $\gamma_{\rm sub} \sim 0.86$.
	(b) The sub-linear relation between street length, $\ell$ and $p^{\gamma_{\rm sub}}$ is shown for the empirical data. It follows
	the theoretical prediction almost perfectly (red line).
	(c) As an example for a super-linear scaling law the relation between city GDP and  $p^{2-\gamma_{\rm sub}}$ is shown.
	Red lines represent the linear regression.
\label{fig::results}}
\end{figure*}

The physical aspect of cities is largely composed by its buildings and its street network.
These street networks can be characterized with a fractal dimension $1<d_i<2$, which
can be directly measured with box counting from maps, see SI.
The estimation of the dimension of the population fractal, $d_p$, is harder to obtain due to limitations in the data.
The 3D information of the population distribution is not directly available. To compute it nevertheless,
the basic idea is to decompose $d_p=d_{p_p}+ \beta$ into a planar (or projected) part, $d_{p_p}$,
that can be directly obtained as the fractal dimension of spatial population data \cite{populationGrid2015},
and a component that captures the ``fractality'' of the vertical component, $\beta$, which can be approximated from
data on the height of buildings \cite{OpenStreetMap}, see Fig.~\ref{fig::fractalPop} (d). For details, consult SI.
Since people live in buildings, and since buildings are aligned along streets, it is natural
to assume that $d_{p_p}$ should be close to $d_i$. In SI Fig.~\ref{fig::dppvsdi} we show empirically that $d_i \sim d_{p_p}$ holds indeed.
Figure~\ref{fig::results} (a) shows  the measured dimensions $d_i$ and $d_{p}$ for 1000 UK cities (every dot is a city) as a function of their population.
Results for other mayor European countries (DE, FR, ES, IT) are shown in SI Fig.~\ref{fig::resultsALL} (a).
Both dimensions exhibit a clear dependence on population size, $p$, therefore their ratio although numerically almost constant
across countries, also shows a small dependence on city size which saturates as cities grow larger. As explained in depth in the SI, this dependence is important, since it means that the exponent cannot be characterised by a single number and it is something that needs to be contemplated in order to avoid contradictions.
The sub-linear scaling exponent, $\gamma_{\rm sub}$, allows us to almost perfectly reproduce the
empirical length of street networks $\ell$, see Fig.~\ref{fig::results} (b).

To obtain the sub-linear exponent we take the following steps.
Given the fractal dimension $d_i$, the total length of the street network scales as $\ell = k_i L^{d_i} $, where $L$ is the linear extension (scale)
of the city, and $k_i$ is a city specific constant, see Fig.~\ref{fig::fractalPop} (a).
Similarly,  the way the population is embedded in 3D space is a fractal with dimension, $d_i<d_p<d_i+1$, meaning, that population
scales with the linear size of the city, as $p = k_p L^{d_p}$, where $k_p$ is a city dependent constant.
Note that we view the population distributed in space as a cloud of points,
where every person is represented as a point, and its location is given by the 3D coordinates of the apartment where the person lives.
Using this in $\ell=k_i L^{d_i}$, we get $\ell \sim p^{{d_i}/{d_p}}$.
We denote the sub-linear scaling exponent by $\gamma_{\rm sub}={d_i}/{d_p}$.
See SI for a more careful derivation.
Note that scaling is tightly related to the definition of the fractal dimension of an object.
To verify how well $\ell \sim k_i L^{d_i} $ and $p \sim k_p L^{d_p}$ are realized empirically, consult SI Fig.~\ref{fig::approximations}.


The origin of the super-linear scaling exponent has been associated with human interaction densities \cite{bettencourt2013origins}.
It has been argued that one reason why cities are liveable and why urbanization continues to increase is because
they facilitate interactions between its inhabitants.
The number of interactions (as approximated by the number of cellphone calls) as a function of city size follows a scaling law with a super-linear scaling exponent that is close to $1.12$ \cite{schlapfer2014scaling}. It was further argued \cite{bettencourt2013origins} that the observed super-linear scaling exponents in urban data can be explained as a consequence of super-linear interactions of people in cities.
Here we will follow this line of thought and formalize this assumption by estimating the number of interactions
as a consequence of the geometry of the street network  and how people are distributed in three dimensions.

To compute the number of interactions that can happen on a unit $1 \times 1$ square, we first recall how the population scales.
Given a 3D grid of cubes of linear size $\epsilon$, the average number of people living in an $\epsilon$-box is
$\langle p\rangle_\epsilon=k_p\epsilon^{d_p}$,
where $k_p=\langle p\rangle_1$ is a city-specific constant, the number of people living in a box of size 1.
It is related to the number of people per square meter that can live in a flat.
If we now choose a box size that contains the entire city ($\epsilon=L$), the population can be expressed as
  \begin{equation}
  p\sim   k_p  L^{d_p}  = \langle p\rangle_1  L^{d_p}\quad ,
  \label{eq::totalPa}
  \end{equation}
where $L$ is the length of a square that contains the  city.
The number of boxes of size $1$ occupied by the population is $L^{d_p}$.
The same argument can be repeated for the projected population.
The average number of people living in a $\epsilon \times \epsilon$ square is
$\langle p_p\rangle_\epsilon=k_{p_p}\epsilon^{d_{p_p}}$,
we can obtain with box-counting $d_{p_p}$, and $k_{p_p}$ (see SI).
Writing the population as a function of its 2D projected version, we get
  \begin{equation}
  p\sim\langle p_p\rangle_1 L^{d_{p_p}} \quad ,
  \label{eq::totalPpa}
  \end{equation}
where  $\langle p_p\rangle_1$ is the average number of people in a square of size $\epsilon=1$.
Using Eqs. (\ref{eq::totalPa}) and (\ref{eq::totalPpa}), we can now write
  \begin{equation}
  \langle p_p\rangle_1=\langle p\rangle_1L^{dp-d_{p_p}} \quad .
  \label{eq::averagePeopleProjected}
  \end{equation}
We can now estimate the number of interactions.
If we have $\langle p_p\rangle_1$ people in a square of size 1, the maximal number of their interactions is
$\langle p_p\rangle_1 (\langle p_p\rangle_1-1) \sim \langle p_p\rangle_1^2$.
The total number of interactions $N$ in a city in a single instant would be that value,
multiplied by the number of locations in which that can happen, which is $L^{d_{p_p}}$.
With Eqs. (\ref{eq::averagePeopleProjected}) and (\ref{eq::totalPa}) we get
  \begin{equation}
  \label{eq::interactions}
  N \sim \langle p_p\rangle_1^2 L^{d_{p_p}} \sim
  L^{2d_p-d_{p_p}}  \sim p^{2-\frac{d_{p_p}}{d_p}} \sim p^{2-\gamma_{\rm sub}} \quad .
  \end{equation}
Here we used our empirical finding that the fractal dimension of the projected population follows the dimension of the
street network, $d_{p_p}\sim d_i$; see SI Fig.~\ref{fig::dppvsdi}.
We identify the scaling exponent obtained from interaction densities as the super-linear exponent, $\gamma_{\rm sup}=2-\gamma_{\rm sub}$.
The addition rule follows from this derivation.
As an example for a known super-linear quantity, we show the actual GDP for UK cites in comparison to the theoretical prediction
in Fig.~\ref{fig::results} c.
City GDP data was obtained from  \cite{eurostatGDPNUTS3} and is described in more detail in the SI.

\begin{figure*}[t]
\centering
\includegraphics[width=0.8\textwidth]{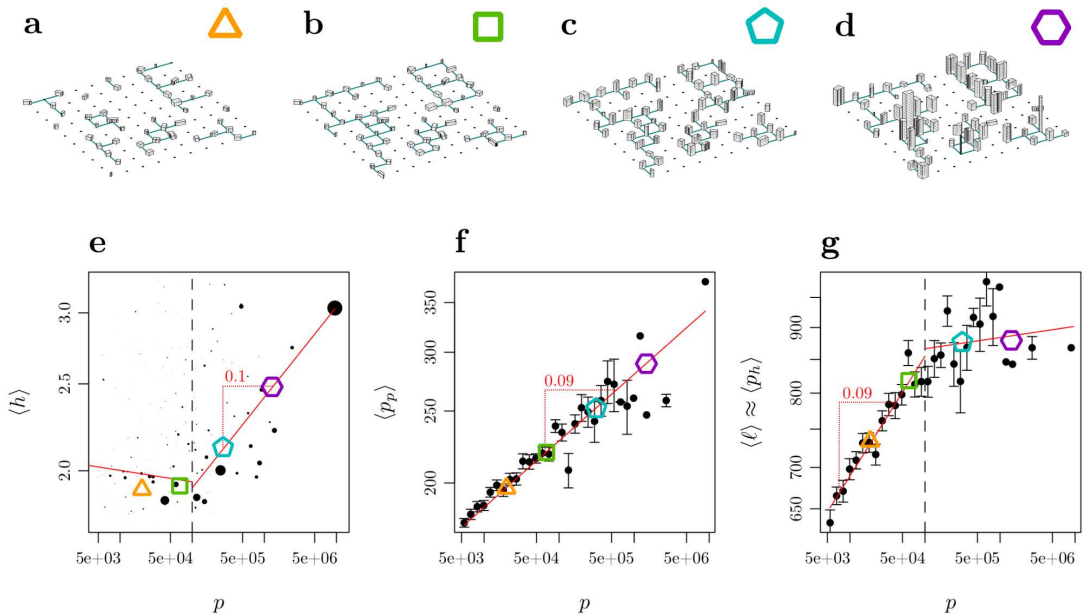}
\caption{Schematic illustration of how a city grows.
	(a)-(b) For low populations the city expands and densifies mostly horizontally, buildings have one or a few levels.
	(c)-(d) From a critical population size upward, buildings begin to grow into the third dimension. In this regime the urban scaling law in the
		average building heights, $\langle h \rangle$, is expected to hold.
	(e) The scaling behavior of the average building height of UK cities, $\langle h \rangle$, is clearly scaling, and follows the theoretical
		prediction for the exponent $1-\gamma_{\rm sub}$. Scaling only appears for populations larger than 100,000.
		The red line indicates the scaling region with a slope of 0.10.
		Below the critical population the growth is marginal.
	(f) The average value of the projected version of the population grows with the same exponent
		$1-\gamma_{\rm sub}$ showing an approximated slope of 0.09. Obviously, there is no critical size, as expected.
	(g) At the critical population level, the average value of the population in each floor, as measured by $\langle p_h \rangle$,
		saturates, and can no longer grow. Up to this point, the street network densifies to absorb the increase of the population.
		Beyond the point, buildings start to grow in height. Notably, this happens with the same exponent, $1-\gamma_{\rm sub}$. In (e) each point is a city and the size of the points represents the number of buildings that are digitised in that city. In (f) and (g) each point is the average for similar sized cities using log-bins.
\label{fig::explanation}}
\end{figure*}


We can now make a prediction about the scaling behavior of the height of buildings, measured in numbers of floor levels, $\langle  h \rangle$.
Assuming that the depth of buildings have a well defined average (buildings have a similar dimension measured from the street to the back of the building),
and that in each building the same number of people live per floor level (because there are physical limitations to how many people can live on a square meter), then, the average population per level, $\langle p_h\rangle_\epsilon$, is proportional to the average length of roads,
$\langle p_h\rangle_\epsilon\sim \langle \ell\rangle_\epsilon$.
Since also $\langle p_h\rangle_\epsilon=\langle p_p\rangle_\epsilon/\langle h\rangle$, we have $\langle p_p\rangle_\epsilon/\langle h\rangle\sim \langle \ell\rangle_\epsilon$ and given that $p\sim\langle p_p\rangle_1 L^{d_{p_p}}$ (Eq.~\ref{eq::totalPpa}), and similarly, $\ell \sim k_i L^{d_{i}}=\langle \ell\rangle_1 L^{d_i}$ using $d_{p_p} = d_i$ we have:

\begin{equation}
	 \frac{p }{\langle h\rangle L^{d_{p_p}}}\sim \frac{\ell}{ L^{d_i}} \text{, therefore, } \langle h\rangle \sim \frac{p}{\ell} \sim p^{1-\gamma_{\rm sub}} \quad .
\end{equation}

The population divided by the length of the street network is proportional to the average number of levels in the city, which is an intuitive result.
%
This is shown in Fig.~\ref{fig::explanation} (e) where we observe that for population sizes below $100,000$ practically no growth in the average number of levels is observed; the city grows horizontally, by a densification process in the 2D plane Fig.~\ref{fig::explanation} (g). From $100,000$ people upwards, cities begin to grow more strongly
into the third dimension, with a  scaling exponent of $1-\gamma_{\rm sub}$, exactly as predicted.  A different approach to study scaling of heights was presented in \cite{schlapfer2015urban}.

We can further derive the exponent for the projected average population.
Since by definition, $\langle p_h\rangle_\epsilon=\langle p_p\rangle_\epsilon/\langle h\rangle$, and because the average population per level,
$\langle p_h\rangle_\epsilon$, is expected to reach saturation at a certain maximum density,
we expect that $\langle p_p\rangle_\epsilon \sim \langle h\rangle \sim p^{1-\gamma_{\rm sub}}$.
This result perfectly matches the empirical situation that is shown in Fig.~\ref{fig::explanation} (f).
Finally, in Fig.~\ref{fig::explanation} (g) we show that there exists a critical point beyond which the average population per level,
$\langle p_h\rangle$, can no longer increase because it saturates. At this point, again for about $10^5$ people,
the number of levels must increase to keep up with the increase of $\langle p_p\rangle$, as shown in Fig.~\ref{fig::explanation} (e). This change of regime is probably related to the critical population that determines the transition from a mono-centric to a poly-centric city \cite{louf2013modeling,louf2014congestion}.
Since the two measurements of $\langle p_p\rangle$ and $\langle h\rangle$ are completely uncorrelated,
and respective data comes from two independent data sources, the soundness of our geometrical
approach can be tested by showing that $\langle p_p\rangle$ and $\langle h\rangle$ grow with the same exponent for cities above a population of $10^5$, which is indeed the case, see Fig.~\ref{fig::explanation} (e) and (f) and in the SI, Fig.~\ref{fig::correspondance}.

We summarize the urban scaling exponents that are explainable within the proposed geometric framework in Tab. \ref{table1}. Since the exponent varies for cities of different sizes, we have used the results for the largest cities of the values $\gamma_{\rm sub}\sim 0.86$ and $d_p\sim 2.14$.

\begin{table}[t]
\caption{Summary of the urban scaling exponents that are explainable within the geometric framework for the results obtained in the UK.
The maximum values of $\gamma_{\rm sub}$ obtained for the different countries are
$\gamma_{\rm sub}^{\rm UK}= 0.86$, $\gamma_{\rm sub}^{\rm FR}= 0.79$, $\gamma_{\rm sub}^{\rm DE}= 0.81$, $\gamma_{\rm sub}^{\rm ES}= 0.82$, $\gamma_{\rm sub}^{\rm IT}= 0.81$.
}
\begin{tabular}{l c c c}
\hline
quantity 	&  theory 	& measured & reference\\
\hline
street length, $\ell$ 				& $\gamma_{\rm sub}$ 		&  0.86  	& here \\
average height, $\langle h \rangle$ 	& $1-\gamma_{\rm sub}$ 		&   0.10  		&  here, Fig.~\ref{fig::explanation}(e)  \\
interactions, $N$ 				& $2-\gamma_{\rm sub}$ 		&  $1.12$ 		& \cite{schlapfer2014scaling} \\
city GDP 		 				& $2-\gamma_{\rm sub}$ 		&  $1.12$	 	&  \cite{bettencourt2016urban} \\
proj. pop., $\langle p_p \rangle$  & $1-\gamma_{\rm sub}$ 	&  $0.09$	 	& here, Fig.~\ref{fig::explanation}(f)  \\
city area&$\frac{2}{d_p}\sim 0.93$&$0.91$&here, (SI) Fig.~\ref{fig::area}\\
\end{tabular}
\label{table1}
\end{table}

\section*{Discussion}

Urban scaling laws are deeply related with the ways humans move, live, act, and interact in a city.
The way these actions can happen is strongly governed and constrained by the specific geometry of a city.
A geometrical measure that is able to capture these constraints is the ratio between the fractal dimension
of the infrastructure (street) network and the fractal that represents how the population is distributed in 3D space.
We find that this geometric ratio characterizes a city much better than the fractal dimension of its streets or the population alone;
it appears naturally in many urban scaling relations.
We claim that urban scaling laws, which emerge as a result of the interplay between the structures where people are located
and the structures they can move on, can be expressed in terms of this ratio.
We explicitly showed that this geometric framework leads to predictions that are in excellent agreement with actual data
for the scaling laws of the length of street networks and heights of buildings.
For the latter, the value of the exponent is determined by how the heights of buildings change,
once cities start to expand into the third dimension, which happens at a critical city size of about 100,000 people.
Furthermore, the geometric ratio explains a number of very different aspects of scaling in a perfectly coherent way.

In summary, a fractal geometry perspective on cities allows us to accomplish the comprehensive understanding
of the origin of sub- and super-linear scaling exponents on the basis of geometry alone,
the tight relation between sub- and super-linear scaling, and finally,
a method to systematically relate the fractal dimensions of geometric objects to the exponents of the observed scaling laws.
With the latter we predicted several scaling relations and verified their existence on data.
We summarize these in Tab. \ref{table1}. These geometrical perspective has also allowed us to calculate for the first time individual exponents for each city which shows that the exponent is not constant and depends on city size.


 Cities exhibit a surprisingly stable geometrical ratio across countries and cultures showing even a similar dependence to population size. The nature of this behaviour is still unknown. We are merely proposing a new viewpoint in this discussion in the hope that it will serve as one more tool to deepen our understanding on urban scaling laws.

%
%
%
\section*{Acknowledgements}
We acknowledge support from FFG project under 857136.
The authors wish to thank Vito D.P. Servedio and Elsa Arcaute for discussions on the subject.
\bibliographystyle{ieeetr}

\pagebreak

\onecolumngrid

\section{Supplementary Information}

\begin{appendix}

\subsection{Describing the dependence of the scaling exponent on the population}
In order to depict and calculate the exponent of the scaling, it is common to use a single number for each system of cities as if the behaviour among cities of different sizes would be constant. After all, the variation is numerically very small and even we have done the same in this work for summarising purposes. But this is not exactly the case and it can lead to contradictions, like exponents being measured as super- or sub-linear in different works.
We wanted to show in this appendix, that a single number cannot describe the true and full behaviour of the quantities studied in this work since $\gamma_{\rm sub}$ is shown to depend on the population size. Up to our knowledge, this work is the first to obtain individual values of the scaling exponent for cities of different sizes, so this issue has not been noticed before in the literature.

\begin{figure}
\centering
\includegraphics[width=1\textwidth]{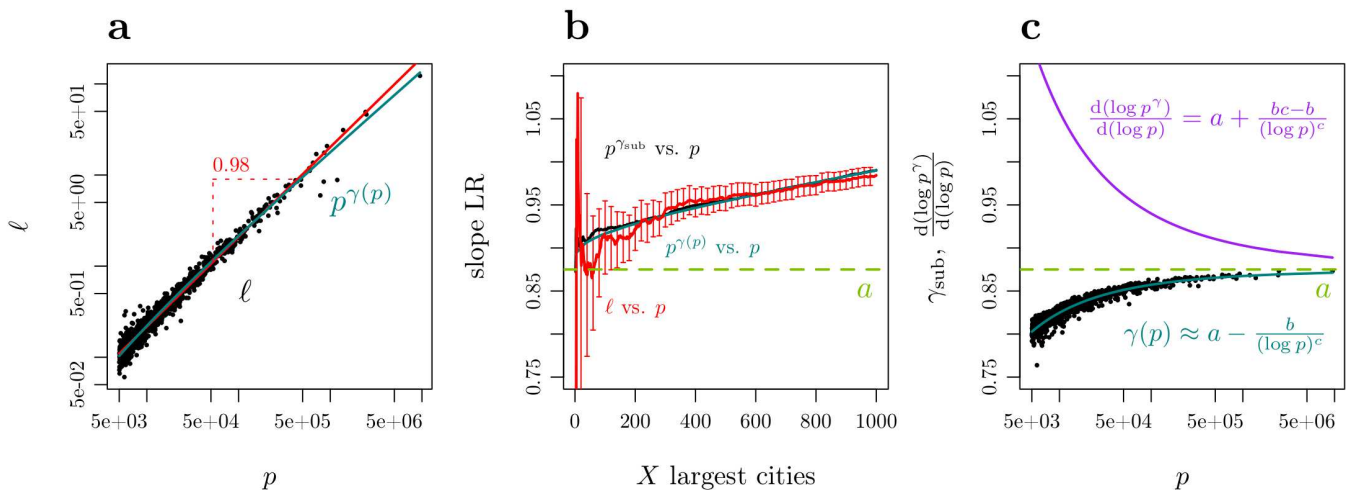}
\caption{Showing the implications of the dependence of $\gamma_{\rm sub}$ on the population. (a) Represents a linear regression (in a log-log scale) for all cities in the UK. (b) Shows the values of the exponents obtained using linear regressions with increasingly larger number of cities between $\ell$ against $p$ (red), $p^{\gamma_{\rm sub}}$ against $p$ (black) and its approximation $p^{a-{b}/{(\log p)^c}}$ against $p$ (blue).  (c) $\gamma_{\rm sub}$ as black points, blue line is its approximation $a-\frac{b}{(\log p)^c}$  and black line is the local slope in a log-log plot of $p^{a-{b}/{(\log p)^c}}$ against $p$.
\label{fig::errorCorrelation}}
\end{figure}

\begin{figure}
\centering
\includegraphics[width=1\textwidth]{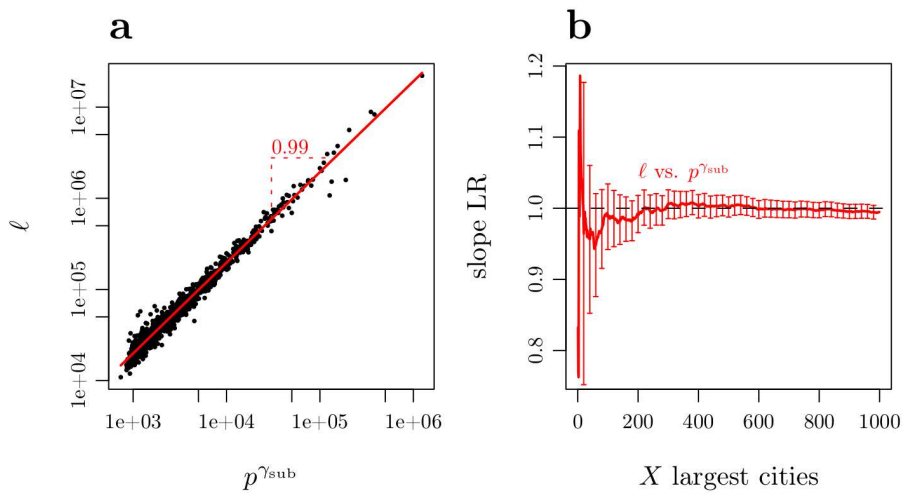}
\caption{Showing how the curvature of $\ell$ is corrected through the calculated $\gamma_{\rm sub}$. (a) linear regression between $\ell$ and $\gamma_{\rm sub}$ for the whole set of cities. (b) Slopes of the linear regressions between $\ell$ and $\gamma_{\rm sub}$ for increasingly larger sets of cities, showing how the relationship is constant and independent of the lower cutoff, given that our calculations have corrected for the curvature of $\ell$.
\label{fig::correctingCurvature}}
\end{figure}

As shown in Fig.~\ref{fig::errorCorrelation} (a), if we take the population and the total length of roads of the UK and perform a linear regression against its population with the full set of 1000 cities to calculate the scaling exponent of the relation we get a value for the scaling law of $0.98$. But, this would mean that the behaviour is practically linear and our measurements show that the maximum value for $\gamma_{\rm sub}$ in the UK is $0.87$ which is a sub-linear exponent. Furthermore, looking at Fig.~\ref{fig::errorCorrelation} (a), we see that the linear regression of $\ell$ vs. $p$ produces this slope of 0.98, while in Fig.~\ref{fig::correctingCurvature} (a), the linear regression of $\ell$ vs. $p^{\gamma_{\rm sub}}$ gives a slope of 0.99. Both are contradictory and incompatible unless we understand that $\gamma_{\rm sub}$ is not a single value, but a function of $p$ Fig.~\ref{fig::errorCorrelation} (c).

What is truly happening, as shown in Fig~\ref{fig::errorCorrelation} (b), is better explained by performing a linear regression of $\ell$ vs. $p$ with increasingly larger sets of cities (starting with the largest city and including at each step smaller cities). This shows a behaviour that depends on the lower cutoff chosen, the more cities we include into our measurement the larger the result that we obtain. In this figure we perform the same calculation for linear regressions between $\ell$ vs. $p$, $p^{\gamma_{\rm sub}}$ vs. $p$ and its approximation $p^{a-{b}/{(\log p)^c}}$ vs. $p$. As we can see, all the 3 curves show a similar behaviour once the number of cities is large enough for the slope not to be spurious (above the largest 100 or 200 cities), and they present a non-constant value that increases the more cities we include in our calculation. This variation of the scaling exponent (considered as a single value) as the lower cutoff gets modified was already reported in \cite{arcaute2015constructing}. This is caused by a curvature in the relationship between $\ell$ and $p$, which, albeit small, distorts the results.

We can further explore this subject by fitting a curve to $\gamma_{\rm sub}(p)\sim a-\frac{b}{(\log p)^c}$, and analysing the local slope of the log-log plot of $p^{\gamma(p)}$ ($\sim \ell$) vs. $p$ by taking its derivative in a log-log setting. We thus obtain $\frac{{\rm d}(\log p^\gamma)}{{\rm d}(\log p)}=a+\frac{bc-b}{(\log p)^c}$. Notice that the minus sign has been transformed into a plus sign, so the slope of the curve does not produce the correct value of $\gamma_{\rm sub}$ at any point, it grows in the opposite direction from $a$ and, therefore, it never coincides with the values that created it (Fig~\ref{fig::errorCorrelation} (c)). This means that the slope of the log-log linear regression of $\ell$ vs. $p$ (Fig~\ref{fig::errorCorrelation} (b)) never coincides with the actual values of the exponent that created it, since, independently of the cutoff chosen, it always stays above the $a$ threshold. This dependence of the scaling exponent as the lower cutoff gets modified is therefore an artifact created by the curvature of $\ell$ with respect to $p$ which is fully captured by our calculation of $\gamma_{\rm sub}$. This means, that studies that try to measure a single value for the whole system of cities, will get varying results depending on the number of cities chosen to calculate the exponent. Moreover, it is unfortunately not possible to obtain the true value of the exponent by analysing the relationship of $\ell$ vs. $p$.

Note that the approximation of the scaling exponent function $\gamma_{\rm sub}(p)\sim a-\frac{b}{(\log p)^c}$ is a typical saturation function. So as cities grow larger they will tend to have more stable values, and in fact for cities above the critical population of 100,000 people, these values start to be in a close range, becoming $a$ as the population approaches infinity. Unfortunately, even if the values are fairly close, this does not avoid the issue that we are mentioning here and we cannot obtain the value of $a$ from the data of $\ell$ vs. $p$ as shown in Fig~\ref{fig::errorCorrelation} (b), where the minimum value possible is around 0.90, which is still above 0.86.

In Fig.~\ref{fig::correctingCurvature} we show how our calculation has corrected for the curvature of $\ell$, presenting a constant value of 1 against $p^{\gamma_{\rm sub}}$, independently of the lower cutoff chosen. Therefore, considering the dependence of $\gamma_{\rm sub}$ on the population is fundamental to understand the phenomenon of scaling and should not be overlooked. A singled value exponent cannot describe the scaling behaviour of cities.

\subsection{Definition of city boundaries}

We base the definition of cities on a set of systematic criteria obtained from population data.
The approach allows us to meaningfully compare cities with each other, and also across countries.

Previous work \cite{Arcaute2016} laid out the basic building blocks for the methodology used in this work.
Cities were shown to correspond to clusters of percolation \cite{Stauffer_Aharony_percolation1994} from the road network at the threshold,
where the average fractal dimension of the system was maximised. The algorithm is similar to the one proposed for the City Clustering Algorithm \cite{rozenfeld2008laws,rozenfeld2011area} with the added particularity of adding a mechanism to determine a specific threshold at which to obtain the cities.

\begin{figure}
\centering
\includegraphics[width=.85\textwidth]{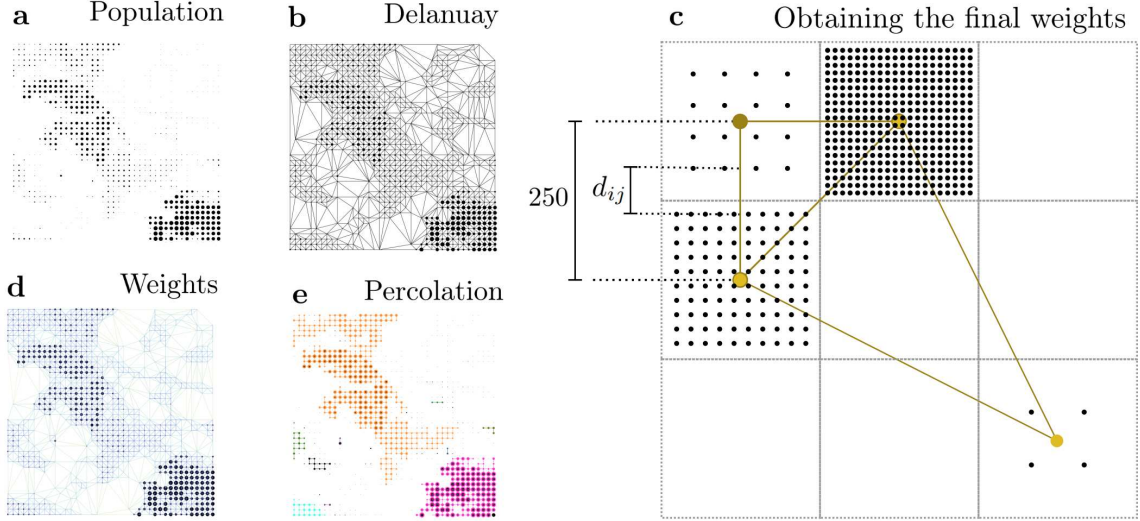}
\caption{Methodology to construct the network of the population in order to obtain the boundaries of cities. (a) population data, (b) delanuay triangulation of the population. (c) creating the fine-grained distances. (d) assigning the weights to the network. (e) calculation the percolation for a threshold.
\label{fig::delanuay}}
\end{figure}

We will be using the population dataset \cite{populationGrid2015} which gives us the population living in a square of $250\times 250$m (Fig~\ref{fig::delanuay}.a). The first step of our methodology is to create a delanuay triangulation (Fig~\ref{fig::delanuay}.b) using the centers of the squares in order to produce a planar network that maintains the adjacency relations. Notice that we cannot use directly a grid connecting the points, because all locations that have 0 population have been removed from the dataset to avoid unnecessary calculations. At this stage the most common weight (distance between nodes) in the network would be 250m, but in order to obtain the final result we need to modify these weights in order to take into account how many people are living into a specific square.

This modification is approached, assuming that the population in a given square is expanding itself as much as possible and in order to make the least amount of assumptions we consider that the spacing between each person will be distributed evenly along the space of the square (Fig~\ref{fig::delanuay}.c). Therefore, the distance from the outermost person living in a square, to the limit of that square would be given by:
\begin{equation}
\label{eq::popDistances}
d_{i}=\frac{250}{n^{0.5}+1}
\end{equation}
where $d_i$ represents this internal distance from the last person living in the square to its limits, and $n$ is the number of people living in the square. After computing these values for the full set of nodes in our network, the final distance between two nodes $i$ and $j$ connected by a link is given by
\begin{equation}
\label{eq::popDistances2}
d_{i,j}=d_{i,j,0}-250+d_i+d_j
\end{equation}
where $d_{i,j}$ is the final distance, $d_{i,j,0}$ is the initial distances between the population centres, $d_i$ is the distance from the outermost person in node $i$ to its limits and similarly $d_j$ for node $j$. We can now assign those values as weights to our network (Fig~\ref{fig::delanuay}.d) and proceed to the last step of our methodology.

\begin{figure}
\centering
\includegraphics[width=1\textwidth]{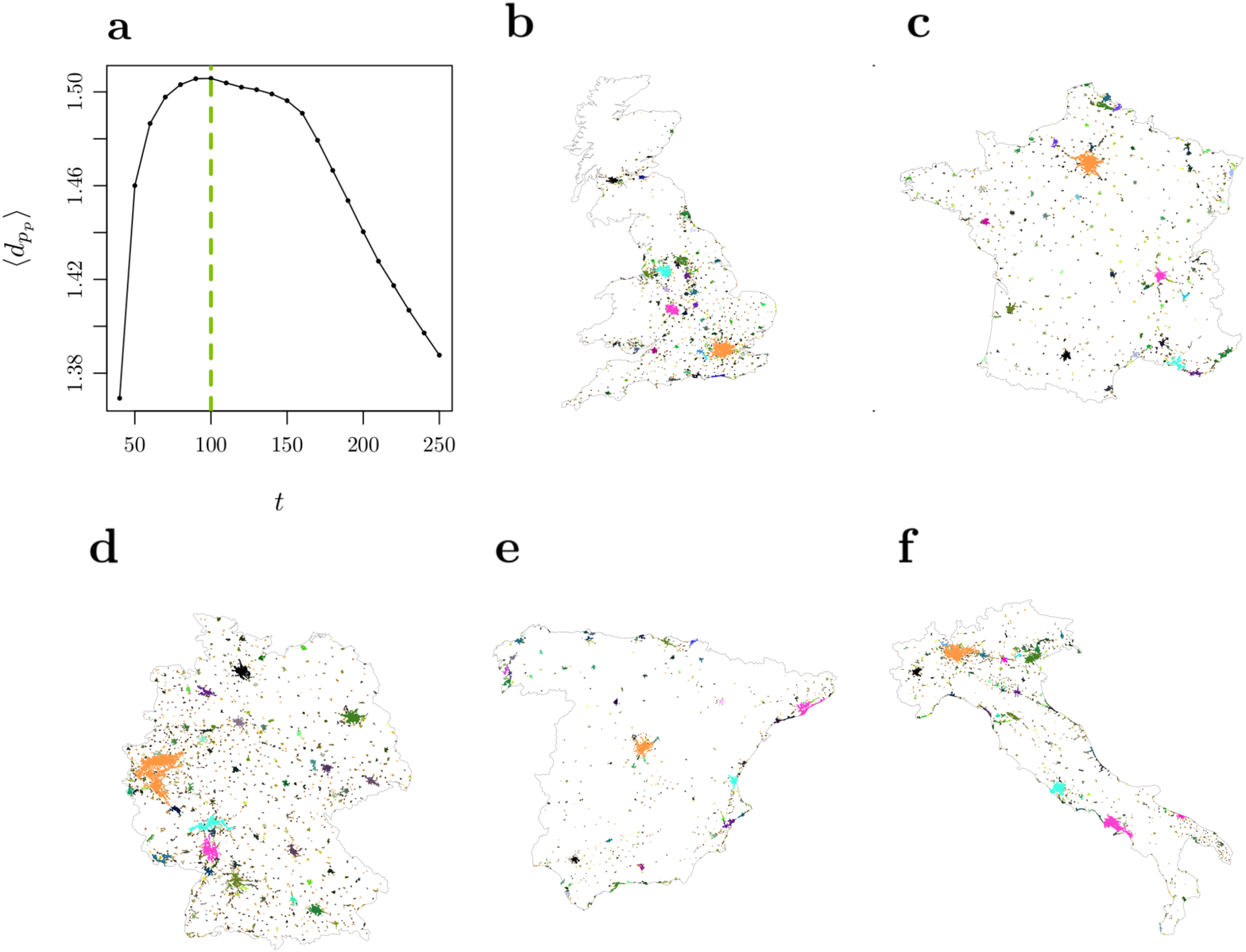}
\caption{Determining the threshold to obtain cities. (a) weighted average of the fractal dimension in FR. (b) final set of cities for UK. (c) final set of cities for FR. (d) final set of cities for DE. (e) final set of cities for ES. (f) final set of cities for IT.
\label{fig::determineThresholdBoundaries}}
\end{figure}

This last step consists on calculating a percolation of that network (Fig~\ref{fig::delanuay}.e) for the whole range of possible thresholds (from the minimum weight to the maximum) and for the set of clusters obtained in each threshold calculate the fractal dimension $d_{p_p}$ for each cluster. We then obtain a weighted average of those values using the logarithm of the population of each city as a weight:
\begin{equation}
\label{eq::popAverFractal}
\langle d_{p_p}\rangle=\frac{\sum_{\forall i} d_{p_p}(i)\log p(i)}{\sum_{\forall i} \log p(i)}
\end{equation}
where $d_{p_p}(i)$ is the fractal dimension of the projected population of city $i$ and $p(i)$ is its population. We define the threshold at which cluster correspond to cities at the maximum of this measure. This is shown in Fig.~\ref{fig::determineThresholdBoundaries}.a, along with the final set of cities obtained for France.


%
%
%
%
%

\subsection{Measuring fractal dimensions: box-counting for the street network}

After obtaining the area that each city occupies we extract its roads from the planet file of Open Street Maps \cite{OpenStreetMap} and calculate the box-counting dimension \cite{falconer2004fractal}. A standard algorithm for calculating this dimension is to calculate the number of boxes that are occupied using different sizes of boxes and the dimension is then defined as:
 \begin{equation}
 \label{eq::fractalDimension}
 d_i= \lim_{\epsilon \to 0}  \frac{\log N(\epsilon) }{\log 1/\epsilon}
 \end{equation}

This happens because as mentioned before in the text, the average number of elements inside a box of size $\epsilon$ follow
\begin{equation}
\label{eq::fd2}
\langle x\rangle=k \epsilon^d
\end{equation}
 which means that $x/N_{\epsilon}=k \epsilon^d$ and therefore $N_{\epsilon}=x/k \epsilon^{-d}$. We use a small modification of this algorithm which produces identical results but at the same time allow us to measure also the multiplying constant $k$. This is based in using directly Eq.~\ref{eq::fd2} and measuring the average number of elements inside a box of size $\epsilon$ and then comparing that average against the size of the boxes used (only for the occupied boxes, the $N$ in the previous equations). Then,
 \begin{equation}
 \label{eq::fd3}
 d_i= \lim_{\epsilon \to 0}  \frac{\log \langle x\rangle_\epsilon }{\log \epsilon}
 \end{equation}

 When measuring real life objects, instead of perfect mathematical fractals, it is impossible to find the fractal dimension for any $\epsilon$ going to 0, and limits must be imposed. Specifically for road networks, we have a very narrow margin. We are using a digital dataset of roads, composed of line objects, and the measure inside each box correspond to the length of roads contained by it. When the size of the boxes go below a certain limit, the relation between the boxes and $\epsilon$ start to curve, eventually leading to a slope of 1, which is the dimension of the lines that compose the digital object. This forced us to set a lower limit of $\epsilon=150m$ for the boxes. Furthermore, fractality in cities is basically produced by the the voids in the urban tissue. This means that it is parks and openings of different size which make that cities do not have a dimension of 2. When boxes are above $\epsilon=500m$ parks start to disappear for the smaller cities and every box is occupied no matter the size, which creates a fake slope of 2.

Values for $d_i$ are shown for cities in the UK in Fig.~\ref{fig::resultsALL} and for four other largest countries in Europe (DE, FR, ES, IT)
in Fig.~SI \ref{fig::resultsALL}.

\subsection{Estimating the fractal dimension of the population}
The basic idea is to decompose $d_p=d_{p_p}+ \beta$ into a planar (or projected) part, $d_{p_p}$,
and a component that captures the fractality $\beta$ of the height of buildings, which can be approximated from
data on the number of levels of the population extracted from \cite{OpenStreetMap}.

\subsubsection{The projected dimension, $d_{p_p}$}
The projected dimension, $d_{p_p}$, can be directly obtained with box-counting as the fractal dimension of the projected spatial population data \cite{populationGrid2015}. In this case we are working with points instead of lines but the algorithm is similar to the one described above. Since the population grid is given every 250m, it is not possible to go below that limit, which makes that this measurement is slightly less precise than the one we obtained for the road network.

\subsubsection{The contribution from building heights,  $\beta$}
The dataset of population that is available is a projection of the real population onto the
two dimensional plane of the surface of the city, therefore we cannot measure its fractal dimension, $d_p$,
and $k_p$ directly. Along the following lines we explain how we can estimate these quantities.

Open Street Maps \cite{OpenStreetMap} is digitalizing three-dimensional information of cities.
This is a work in progress and some countries (such as the UK) are more complete than others.
We can obtain for each city the average number of levels in a building, $\langle h\rangle$,
and the maximum number of levels, $h_m$, as well as how many buildings were digitised in that city. Given this data,
we obtain the average population per level in a $\epsilon\times \epsilon$ square,
$\langle p_h\rangle_\epsilon=\langle p_p\rangle_\epsilon/\langle h\rangle$.

To compute $d_p$ and $k_p$ with box-counting, we need the average number of people in 3 dimensional
boxes of different sizes $\epsilon$. Technically, box-counting needs at least 2 different sizes of boxes, which is,
of course, an extremely bad approximation. However, from the data we only know reliably the average population
per box at two specific $\epsilon$ values. Assuming that the typical floor is 3m high, then for $\epsilon=3$m,
one box fits into every level, and $\langle p\rangle_3=\langle p_h\rangle_3=\frac{k_{p_p}}{\langle h\rangle} \ 3^{d_{p_p}}$.
The second box size is the maximum height of the city, $\epsilon=3 h_{\rm m}$.
The population in each box will be equal to its projected version,
$\langle p\rangle_{3 h_{\rm m}}=\langle p_p\rangle_{3 h_{\rm m}}=k_{p_p}(3 h_{\rm m})^{d_{p_p}}$.
With these two values we can approximate
\begin{equation}
d_p \sim \frac{\log \langle p\rangle_{3 h_{\rm m}} - \log \langle p\rangle_3 }{\log 3 h_{\rm m}-\log 3}
=d_{p_p}+\frac{\log \langle h\rangle}{\log h_{\rm m}} \quad ,
\label{eq::dp}
\end{equation}
Obviously, as argued in the main text, the fractal dimension of the population is the fractal dimension of the planar projection,
plus the fractal dimension of the vertical component. We obtain $k_p$ by first calculating the number of boxes with side length
$\epsilon=3$m ($N_{p,3}$), that is, how many squares there are in 2D ($N_{l,3}$), multiplied by the average height,
$N_{p,3}= \langle h\rangle p/ \langle p_p\rangle_3  =p \langle h\rangle/(k_{p_p}3^{d_{p_p}}) $, and
\begin{equation}
k_p=\frac{p}{N_{p,3}}3^{-d_p}=\frac{k_{p_p}}{\langle h\rangle}3^{d_{p_p}-d_p}
=\frac{k_{p_p}}{\langle h\rangle}3^{-\frac{\log\langle h\rangle}{\log h_{\rm m}}} \quad ,
\label{eq::kp}
\end{equation}

\subsubsection{The relation between  $d_{p_p}$ and $d_i$}
\begin{figure}
\centering
\includegraphics[width=1\textwidth]{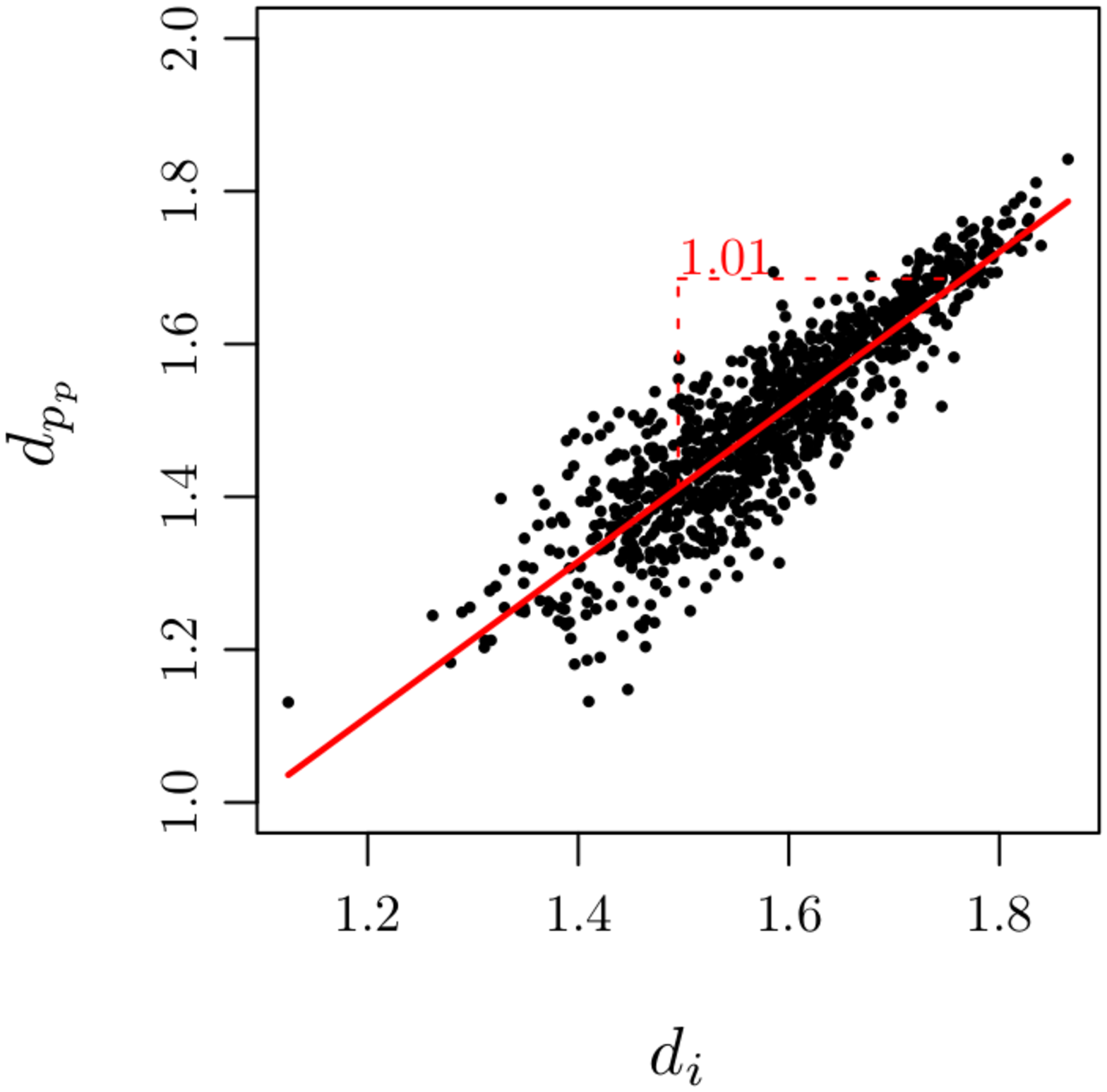}
\caption{Fractal dimension of the population, $d_{p_p}$, versus the dimension of the street network, $d_i$.
It is seen that the two dimensions practically align. The red line represents the linear regression.
\label{fig::dppvsdi}}
\end{figure}

Since people live and in buildings, and those buildings are aligned along streets, it is natural to
to assume that the value of $d_{p_p}$ should be close to $d_i$.
To see that this is indeed the case, see Fig.~\ref{fig::dppvsdi}, which shows the case of the UK.
The linear regression yields $d_{p_p} = 1.01 d_{i} - 0.10$.
It is seen that $d_{i}$ slightly overestimates $d_{p_p}$, which might be due to the lower definition of the population data.
Note that both dimensions are derived from independent and different data sets;
$d_{p_p}$ comes from population data, while $d_i$ is extracted from maps.

\subsection{Derivation of the relation between $\ell$ and $p$ with proportionality factors}

Given squares of length $\epsilon$, the average length of the street network inside each square can be written as
  \begin{equation}
  \langle \ell \rangle_\epsilon=k_i\epsilon^{d_i} \quad.
  \label{eq::averageL}
  \end{equation}
In the same spirit,  we can view the population distributed in space as a cloud of points,
where every person is a point and its location in three dimensions is the apartment where the person lives.
Therefore, given a 3D grid formed of cubes, the average number of people living in a box of size $\epsilon$ is
  \begin{equation}
  \langle p\rangle_\epsilon=k_p\epsilon^{d_p} \quad,
  \label{eq::averageP}
  \end{equation}
where $d_p$ is the fractal dimension of the population and $k_p$ is a city-specific constant.


Combining Eqs.~(\ref{eq::averageL}) and (\ref{eq::averageP})  we can express the
length of the street network as a function of the population living in a box of side $\epsilon$
\begin{equation}
\langle \ell \rangle_\epsilon=k_i\left(\frac{\langle p\rangle_\epsilon}{k_p}\right)^{\frac{d_i}{d_p}} \quad .
\label{eq::averageLasFunctionP}
\end{equation}

The total length of the street network is $\ell = N_{ \ell,\epsilon} \langle \ell \rangle_\epsilon$,
where $N_{\ell,\epsilon}$ is the number of $\epsilon$-sized squares that are occupied throughout the city.
Similarly, the total population is given by $p=N_{p,\epsilon} \langle p\rangle_\epsilon$ with
$N_{p,\epsilon}$ being the number of occupied boxes of size $\epsilon$ that cover the whole three dimensional city.
Using these expressions we get
\begin{equation}
\label{eq::LasFunctionP}
\ell=N_{\ell,\epsilon}k_i\left(\frac{p}{N_{p,\epsilon}k_p}\right)^{\frac{d_i}{d_p}}={k_iN_{\ell,\epsilon}}{(k_pN_{p,\epsilon})^{-\frac{d_i}{d_p}}}{ p}^{\frac{d_i}{d_p}} \sim  p^{\frac{d_i}{d_p}} =  p^{\gamma_{\rm sub}} \quad  ,
\end{equation}
where $\gamma_{\rm sub}$ the sub-linear exponent of the scaling law that relates the length of the road network to the population.

\subsection{Validity of the approximations}

\begin{figure}[t]
\centering
\includegraphics[width=1\textwidth]{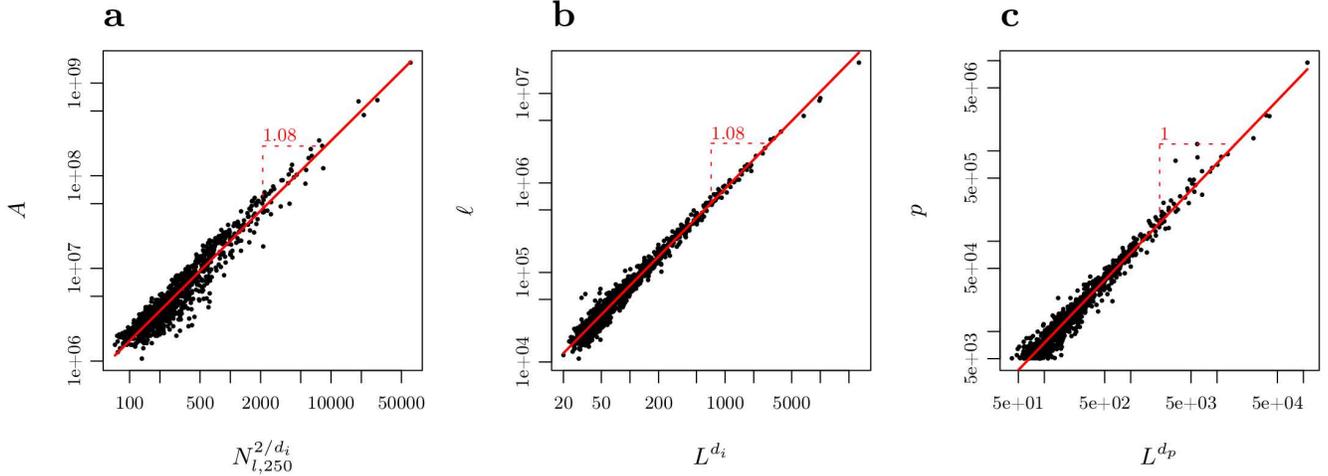}
\caption{(a) Approximation of the area with $N_{l,250}^{2/d_i}$. (b) The relation between $\ell$ and its approximated version using the length of the city $L^{d_i}$. (c) Similar approximation between $p$ and $L^{d_p}$. Green lines denote a perfect relation with a slope of 1.
\label{fig::approximations}
}
\end{figure}

We show in Fig.~\ref{fig::approximations} (b) how the approximated values to $\ell$ and $p$ as a function of the city length scale behave, showing that even though it remains an approximation the values are quite close to the real measurement. $L$ is the theoretical linear length scale of the city. Given the complexity of the measurement that we are taking and that the area of a city has a large amount of noise, we first show that we can predict the area of cities (Fig.~\ref{fig::approximations} (a)) from the number of boxes occupied by the road network at an $\epsilon=250m$, with the following equation $A\sim N_{l,250}^{2/d_i}$ so therefore we will use as our lattice size $L=N_{l,250}^{1/d_i}$ which is a more related measure and shows a better behaviour. Let the reader note that this approximation is never actually used in any calculation performed in the paper, and showing this approximation just remarks what has already been shown many times in other works, that given a lattice size, the number of elements in a fractal is $x\sim L^d$ (see \cite{christensen2002percolation} Fig.~1.10, page 24). The same holds true for the population, $p\sim L^{d_p}$, shown in Fig.~\ref{fig::approximations} (c).

%

\subsection{Results for FR, DE, ES, IT}

\begin{figure}[t]
\centering
France\\
\includegraphics[width=0.7\textwidth]{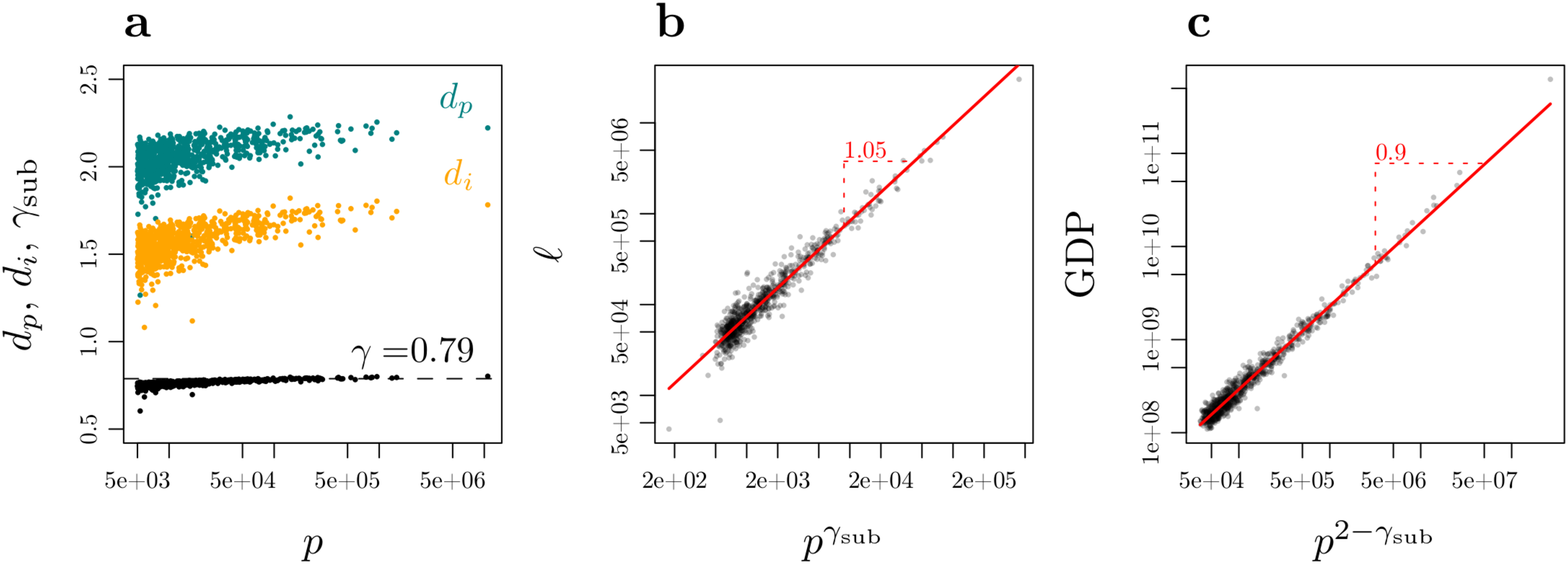}\\
Germany\\
\includegraphics[width=0.7\textwidth]{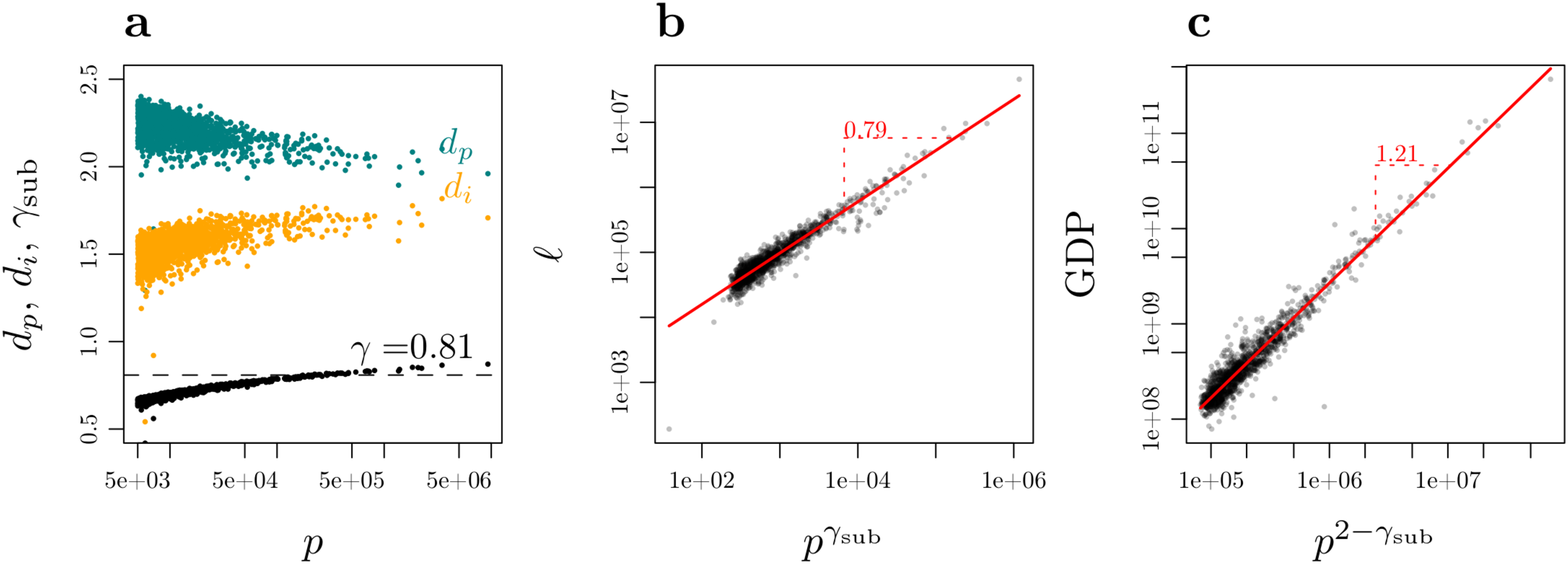}\\
Spain\\
\includegraphics[width=0.7\textwidth]{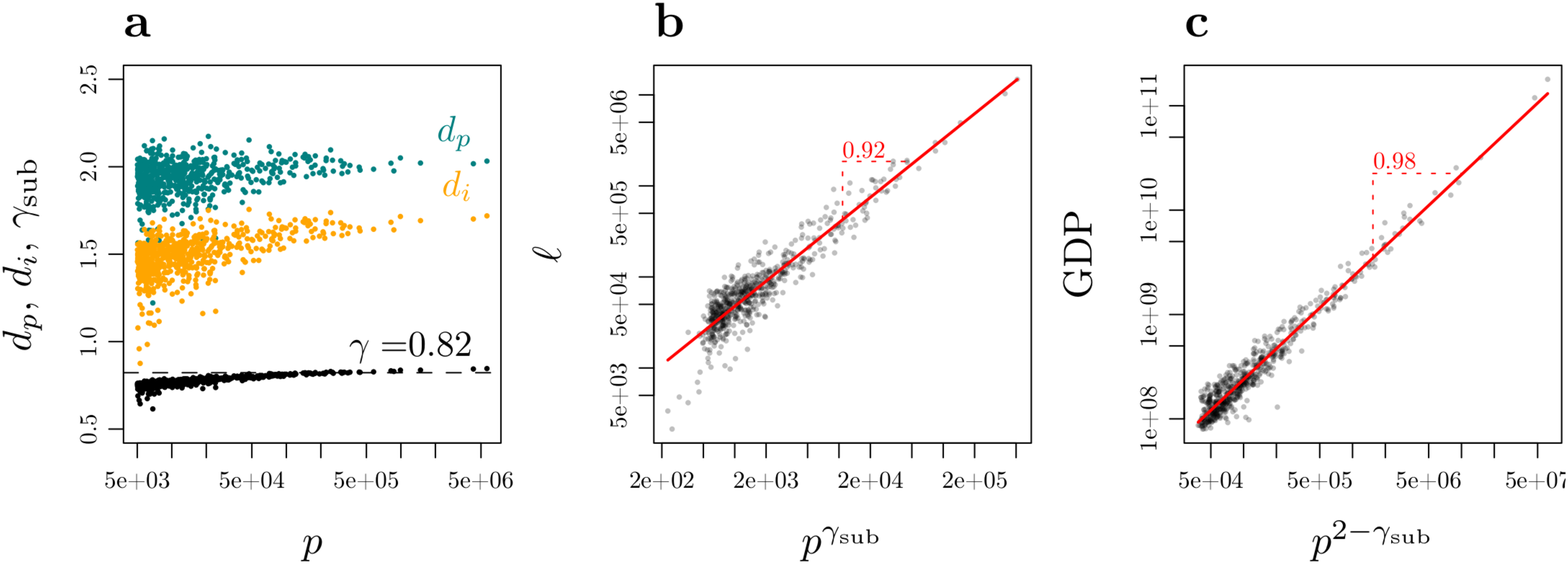}\\
Italy\\
\includegraphics[width=0.7\textwidth]{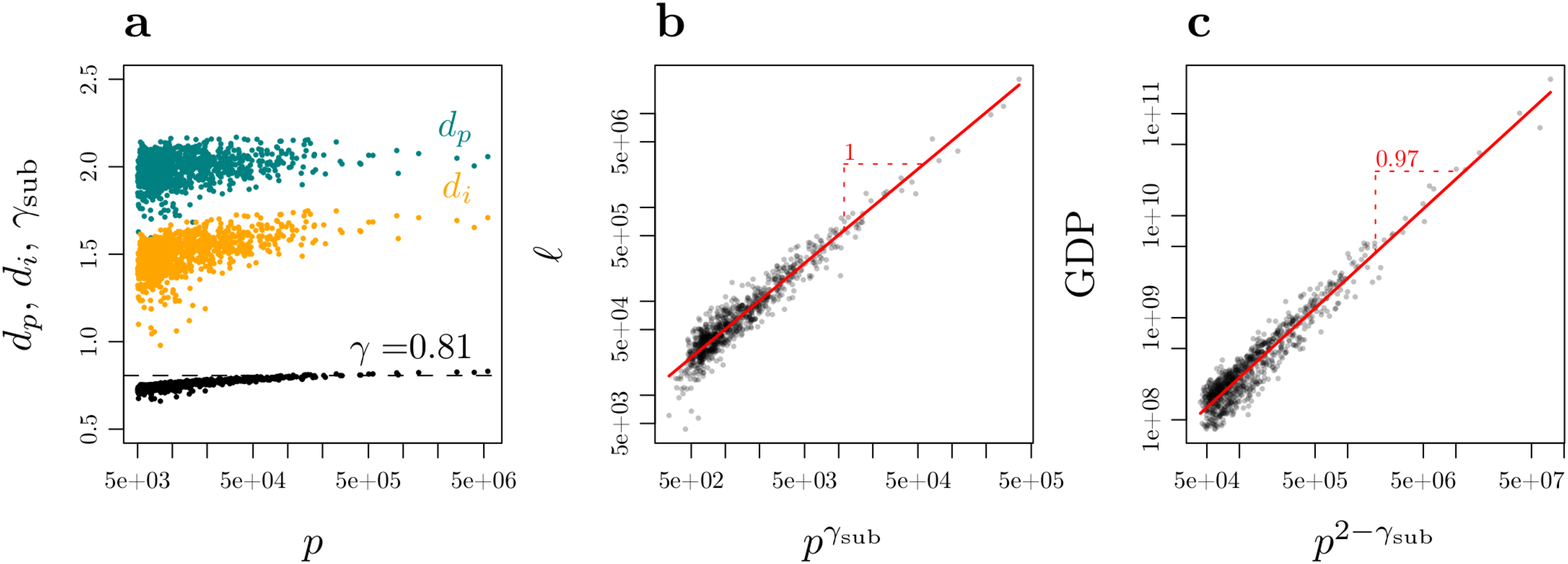}
\caption{Fractal dimensions of the street network, $d_i$, (orange) and for the population, $d_p$ (blue),
		for cities in France, Germany, Spain, and Italy.
		Same setting as in Fig.~\ref{fig::explanation} in the main text, where the situation for the UK is shown.
	(b) The sub-linear relation between street length, $\ell$ and $p^{\gamma_{\rm sub}}$ holds also for the other countries.
		It follows the theoretical prediction (red line).
	(c) Relation between city GDP and  $p^{2-\gamma_{\rm sub}}$. Red lines represent linear regressions.
\label{fig::resultsALL}
}
\end{figure}

We show in Fig.~\ref{fig::resultsALL} the fractal dimensions for the street network, $d_i$, and the population, $d_p$, as measured with box-counting and the
estimation described above for
France (FR), Germany (DE), Spain (ES), and Italy (IT).
Germany presents such bad averages of its heights that it shows an artificial downwards slope of $d_p$. We expect this issue to be solved when more data is available. This is not only caused by the averages of the heights but also because the automated definition of cities determines that the largest city in Germany is the Ruhr region, which is a polycentric urban area composed of several sub-cities. This area still maintains a building stock which corresponds to the size of the individual cities that compose it. This drags the slope of the power-law we use to average the number of levels, forcing it to a negative slope with respect to the population and thus creating this artifact. It can be expected that in a future when the area settles and starts functioning like a single city, its average height will increase.

\subsection{Correspondance between $\langle h\rangle$ and $\langle p_p\rangle$}

\begin{figure}
\centering
\includegraphics[width=1\textwidth]{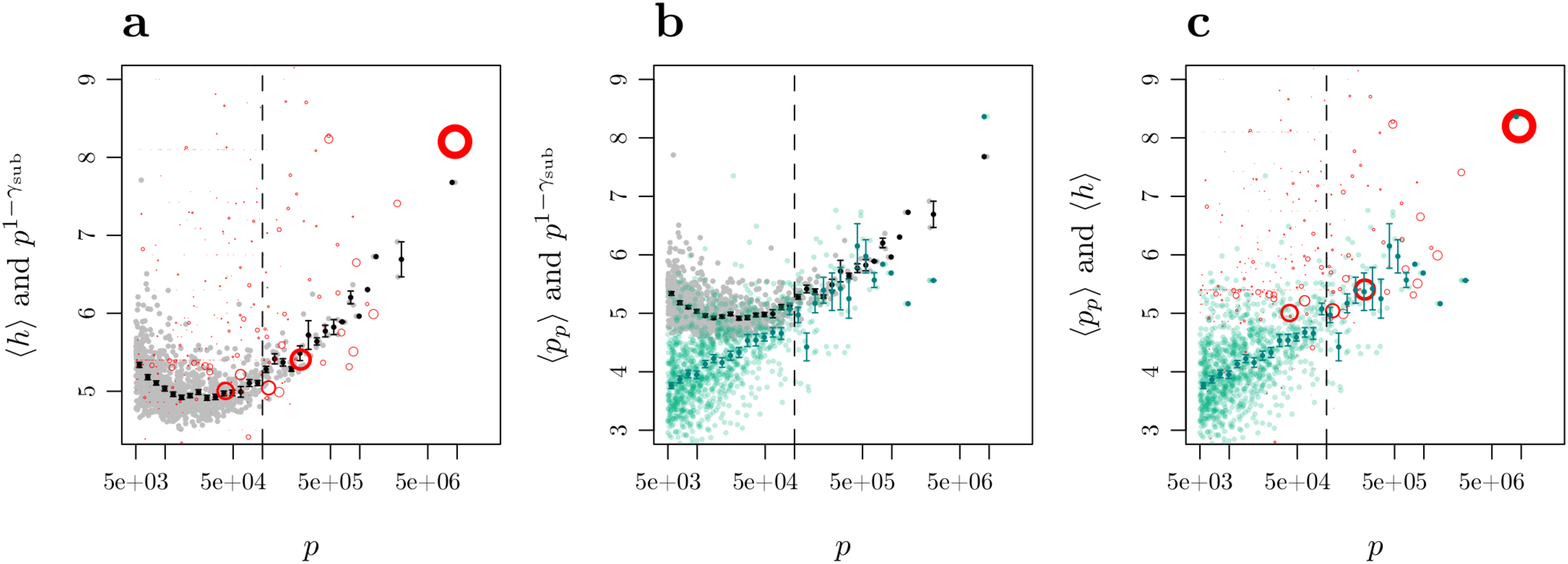}
\caption{Correspondance between $\langle h\rangle$ and $\langle p_p\rangle$ and $p^{1-\gamma_{\rm sub}}$. (a) $p^{1-\gamma_{\rm sub}}$ in black (grey are the true values, black points represent its local averages), and $\langle h\rangle$ where the size and thickness represent the number of buildings. (b) $p^{1-\gamma_{\rm sub}}$ in black (grey are the true values, black points represent its local averages), and $\langle p_p\rangle$ in blue (light blue is true values, while the darker blue shows the averages). (c) Similarly, $\langle h\rangle$ and $\langle p_p\rangle$.
\label{fig::correspondance}}
\end{figure}

We show in Fig.~\ref{fig::correspondance} how the the average number of levels in cities $\langle h\rangle$, its projected population $\langle p_p\rangle$ both follow the same behaviour which is characterised by $p^{1-\gamma_{\rm sub}}$.

\subsection{Datasets}

\begin{figure}[t]
\centering
\includegraphics[width=1\textwidth]{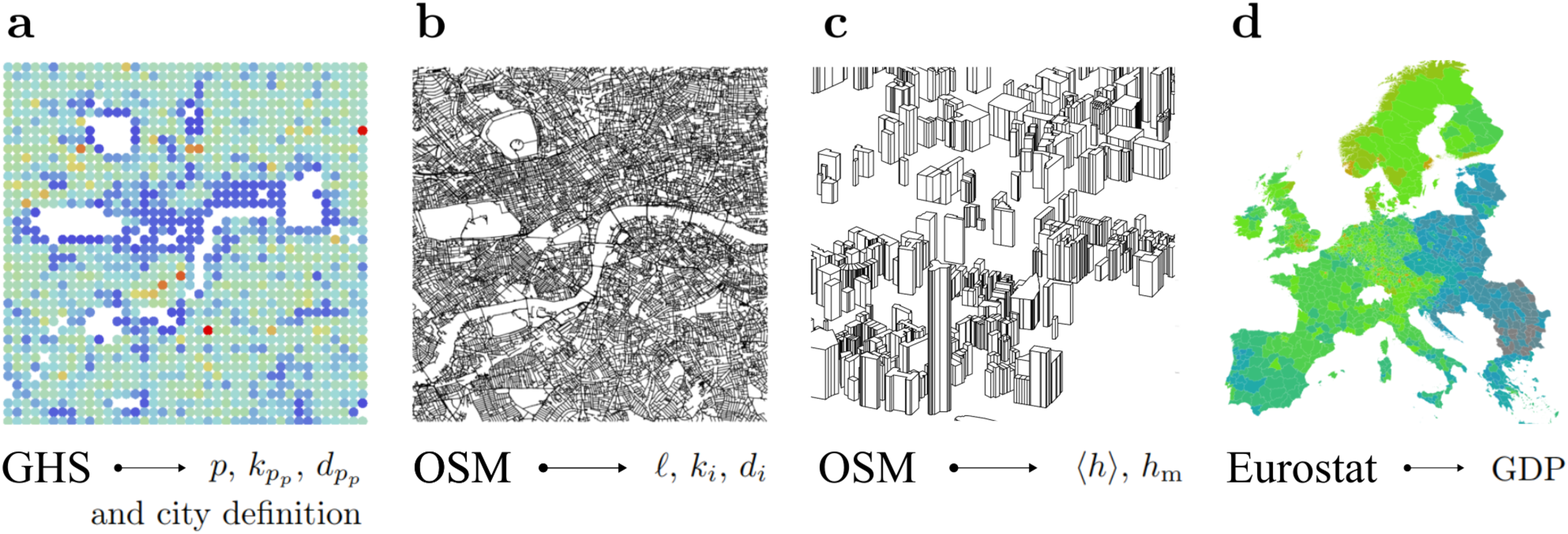}
\caption{The data sources used in this work include,
	(a) population data from the Global Human Settlement,
	(b) road network from OSM,
	(c) height data of buildings from OSM, and
	(d) GDP data per capita from Eurostat at NUTS3 level.
\label{fig::dataSources}}
\end{figure}

\subsubsection{Population data}

We use the Global Human Settlement (GHS) Population Grid from 2015 \cite{populationGrid2015},
which is a global map of the population with a resolution of $250\times250$m,
produced by the European Commission, see Fig.~\ref{fig::dataSources} (a).
From this dataset we obtain the city boundaries using an algorithm based on a percolation
approach in the spirit of \cite{rozenfeld2008laws,Arcaute2016}; see in this SI for more details.
Once the boundaries are determined, we obtain the road networks from the Open Street Map (OSM) dataset
\cite{OpenStreetMap}, Fig.~\ref{fig::dataSources} (b), to measure the total length of the street network, $\ell$,
and the fractal dimension of the infrastructure using box-counting
($d_i$ and $k_i$) for every of the 4,750 European cities in France, UK, Germany, Spain, and Italy.
Only cities above 5,000 people were taken into consideration for the calculations.

\subsubsection{Building heights}

The OSM dataset also contains information about the heights of buildings, in particular the
number of levels in each building Fig.~\ref{fig::dataSources} (c), however data quality is low.
What can be estimated from it in a reliable way is the average number of levels, $\langle h\rangle$,
and the maximum number of levels, $h_{\rm m}$, in every city.

\begin{figure}
\centering
\includegraphics[width=1\textwidth]{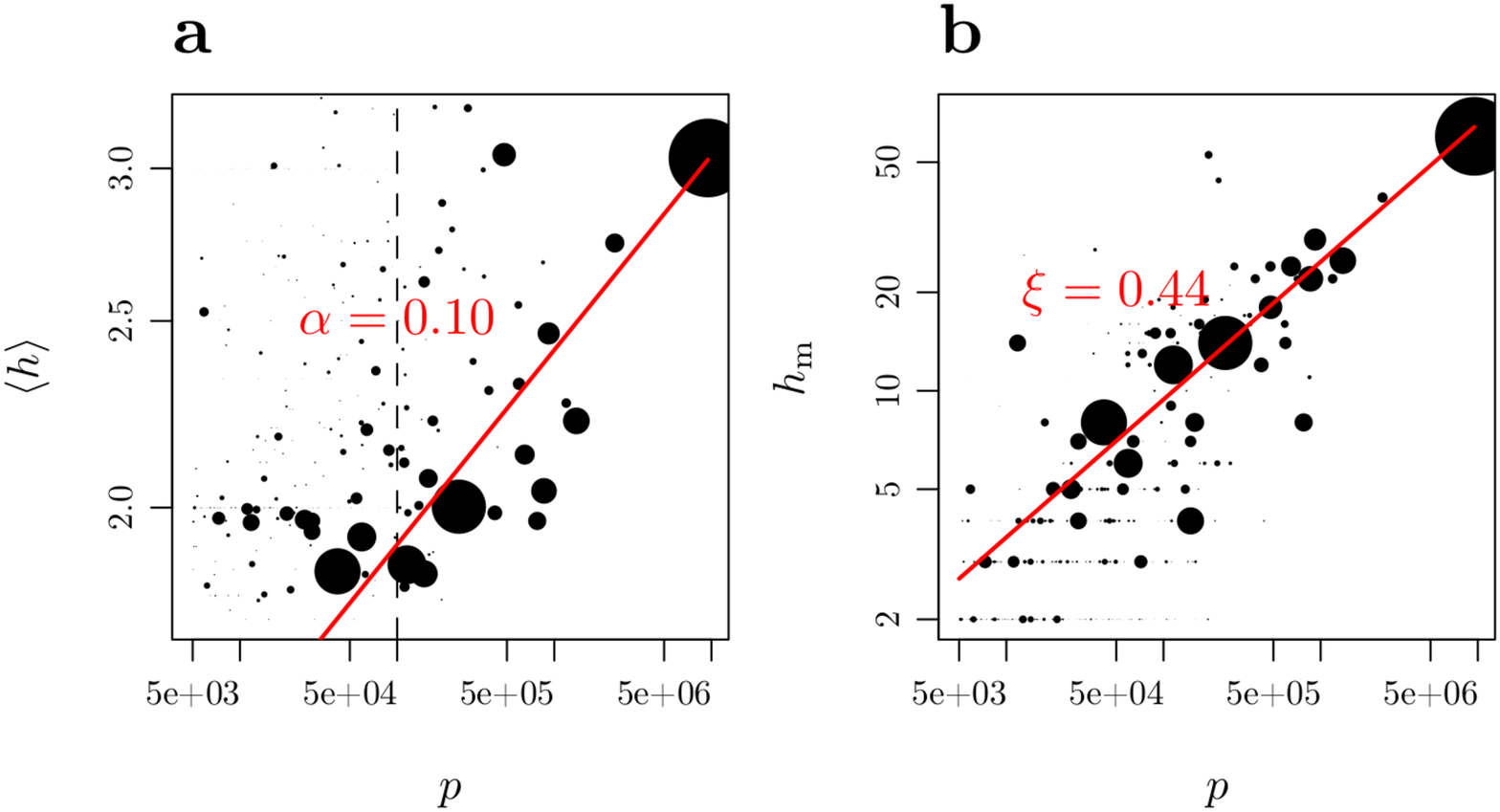}
\caption{Substituting the number of levels by power-laws, data from the UK. (a) Average number of levels. (b) Maximum number of levels. Each circle represents a different city. The area of the circles is proportional to the number of buildings that are already existing in the dataset for each specific city. and red lines represent the power-law that we use to approximate the number of levels.
\label{fig::correlationHeights}}
\end{figure}

Given the low quality of the data relating to the heights of buildings, we substituted the average
number of levels, $\langle h\rangle$, of each city and the maximum heights, $h_{\rm m}$, with
power-laws with respect to $p$ by fitting the two variables.
In order to do this we kept only cities that had at least 10 buildings, and then performed weighted linear regressions using the number of buildings as the weight of each city.
Considering that the average height of buildings saturate in the smaller cities (at some point there is a minimum height), we only perform the correlation to obtain the average number of levels, with cities above 100,000 people (where heights start to increase). Instead the maximum height does show a power law behaviour even in that range so we use the full set of cities.
For these we find that a power law is an approximate good fit, and you obtain the final approximation according to $\langle h\rangle = \langle h\rangle_0 p ^{\alpha}$. The same is done for
$h_{\rm m}$.
The quality of these fits are shown in Fig.~\ref{fig::correlationHeights} where the size of the points represent the number of buildings in each city and each point is a city.

In OSM not all buildings have already been digitized, and the specific values reported should not be taken as final.
Progress differs across countries. The UK has the most buildings and in a large number of cities. This is why we show the
UK results in the main section of the paper.
Note there might be a source for a possible bias in the data that originates from psychological factors on the part of the collaborators of OSM.
It is easier to gain recognition if you digitise ``important'' buildings in the most important cities.
So, more people will tend to pay attention to creating repeated estimates for the empire state building in NY, rather than estimating hundreds of 2 story buildings in unknown regions. This might mean that the dataset is skewed towards higher average heights.
Further improvements in the data might slightly alter the results shown in this paper.

%
%

\subsubsection{Economic data}

To compile the values of the GDP for every city, we use the Eurostat data for the GDP per capita \cite{eurostatGDPNUTS3}
at the NUTS3 level, see Fig.~\ref{fig::dataSources} (d).
Together with the Global Human Settlement population data we calculate the total GDP for every city. This is done by intersecting the polygons in the NUTS3 level with our definition of cities and calculating for each intersection the number of people in that location multiplied by the GDP per capita and finally summing it up for each city.

\subsection{Relation between area and population}

\begin{figure}
\centering
\includegraphics[width=1\textwidth]{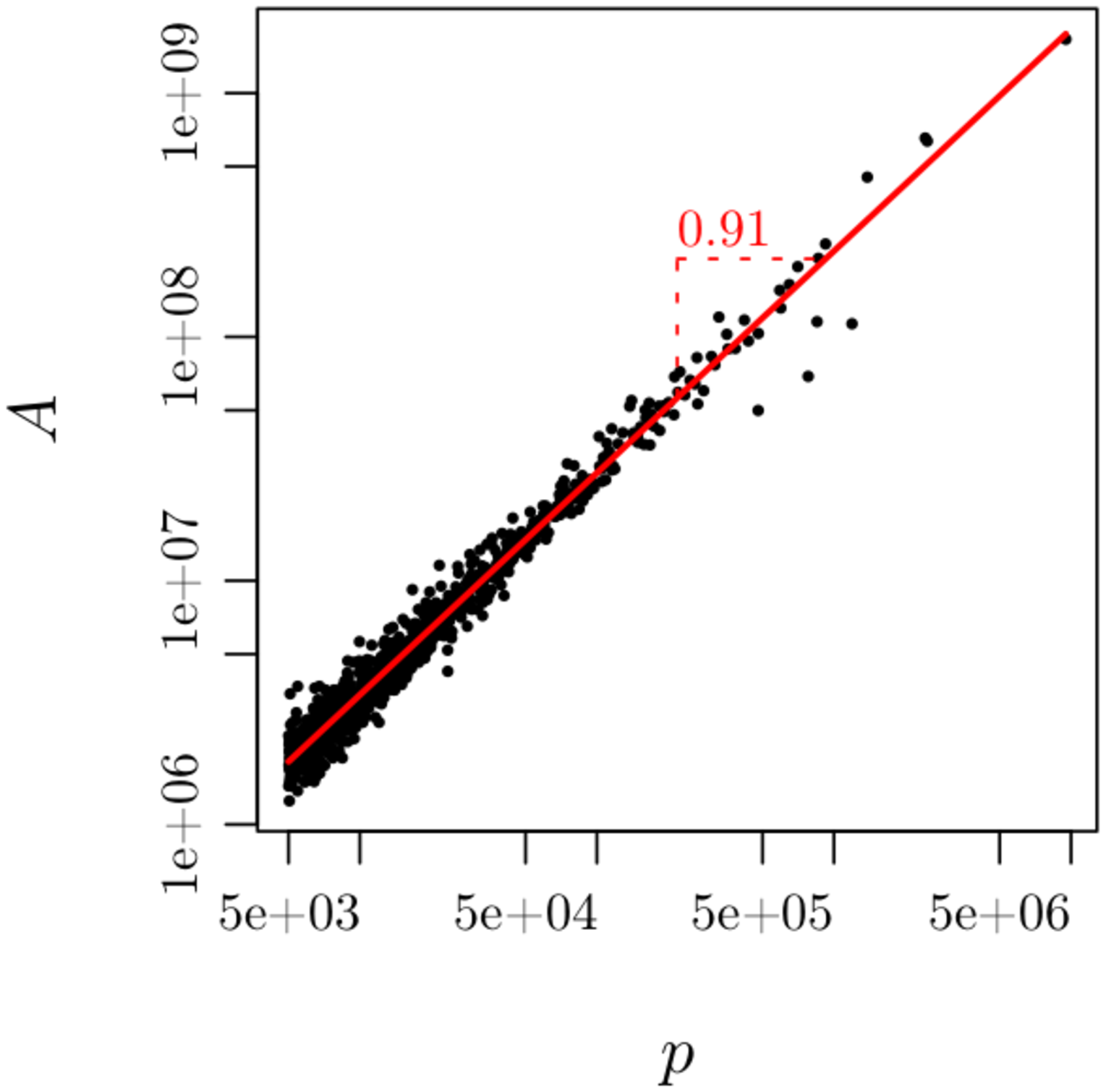}
\caption{Relationship between the area and the population
\label{fig::area}}
\end{figure}

In Fig.~\ref{fig::area} we show the measurement of the relationship between the area and the population.

\end{appendix}
\end{document}